# TWO TYPES OF SPURIOUS DAMPING FORCES POTENTIALLY MODELED IN NUMERICAL SEISMIC NONLINEAR RESPONSE HISTORY ANALYSIS

P. Jehel[(1)], P. Léger[(2)]

[(1)] *Assistant Professor, Laboratoire MSSMat, CNRS-UMR8579, CentraleSupélec, pierre.jehel@centralesupelec.fr*
[(2)] *Full Professor, Department of Civil, Geological and Mining Engineering, Polytechnique Montréal, pierre.leger@polymtl.ca*

## *Abstract*

The purpose of this paper is to provide practitioners with further insight into spurious damping forces that can be generated in nonlinear seismic response history analyses (RHA). The term ‘spurious’ is used to refer to damping forces that are not present in an elastic system and appear as nonlinearities develop: such damping forces are not necessarily intended and appear as a result of modifications in the structural properties as it yields or damages due to the seismic action. In this paper, two types of spurious damping forces are characterized. Each type has often been treated separately in the literature, but each has been qualified as ‘spurious’, somehow blurring their differences. Consequently, in an effort to clarify the consequences of choosing a particular viscous damping model for nonlinear RHA, this paper shows that damping models that avoid spurious damping forces of one type do not necessarily avoid damping forces of the other type.

## 1. Introduction

During severe ground motion caused by an earthquake, the seismic energy imparted to the structures has to be dissipated. A structure collapses if it does not have the capacity to dissipate it. Otherwise, it eventually gets back to rest usually in a damaged state. Damping can be defined as the overall phenomenon that causes energy dissipation in response history analysis (RHA) of a structure subjected to dynamic loading. Numerous mechanisms contribute to damping and need to be modeled to predict the level of structural damage in a post-earthquake state. This damage can then be communicated in terms of structural performance for post-disaster planning.

In computational earthquake engineering, energy dissipation results on the one hand from modeling the nonlinear hysteretic response of the structural system. On the other hand, it results from additional damping that accounts for energy dissipative mechanisms not otherwise represented in the hysteretic model for the structural system. In seismic structural RHA, practitioners have widely been using Rayleigh damping as the model for additional damping. The consequences of this choice have been thoroughly investigated in the past and it is now well known that it can lead to unrealistic (spurious) damping forces.

The equations of motions of a multi degrees-of-freedom (MDOF) structural system with viscous damping and DOFs without associated mass can be written as

$$\begin{pmatrix} \mathbf{M}_{pp} & \mathbf{0} \\ \mathbf{0} & \mathbf{0} \end{pmatrix} \begin{pmatrix} \ddot{\mathbf{u}}_p \\ \ddot{\mathbf{u}}_s \end{pmatrix} + \begin{pmatrix} \mathbf{C}_{pp} & \mathbf{C}_{ps} \\ \mathbf{C}_{sp} & \mathbf{C}_{ss} \end{pmatrix} \begin{pmatrix} \dot{\mathbf{u}}_p \\ \dot{\mathbf{u}}_s \end{pmatrix} + \begin{pmatrix} \mathbf{f}_p^r \\ \mathbf{f}_s^r \end{pmatrix} = \begin{pmatrix} \mathbf{f}_p^e \\ \mathbf{0} \end{pmatrix} \tag{1}$$

where the mass and damping matrices (**M** and **C**) along with the resisting and external forces vectors ($\mathbf{f}^r$ and $\mathbf{f}^e$) and the displacements vector (**u**) have been split into parts pertaining to the *P* DOFs with mass and the *S* massless DOFs. An increase in the resisting forces at time *t* can be written as $\Delta\mathbf{f}^r(t) = \mathbf{K}(t)\Delta\mathbf{u}(t)$ where $\mathbf{K}(t)$ is the tangent stiffness matrix. As the structure is elastic we have $\mathbf{K}(t) = \mathbf{K}_0$. A dot over a quantity indicates derivative with respect to time.

Damping forces $\mathbf{f}^d = \mathbf{C}d\mathbf{u}/dt$ are added to account for energy dissipation sources not otherwise already represented in the structural model. There are many reasons why damping forces are added in seismic nonlinear response history analysis. From a numerical point of view, damping models can be designed so that they damp out non-realistic phenomena coming from high frequencies generated by a small time step in the algorithm [1]. From a physical point of view, structural models rarely explicitly account for all potential energy dissipation mechanism: there are for instance non-structural elements that are generally absent in the model. Potentially, there are numerous energy dissipation mechanisms that can be activated at different scales during the seismic motion history, as illustrated in Fig. 1 for reinforced concrete (RC) buildings. This makes it challenging to develop reliable structural models with explicit representation of all these mechanisms. Accordingly, these numerous mechanisms are all considered together in hysteretic force-displacement laws at the structural element section level, or at the material level in fiber elements.

In elastic analysis, an efficient method for solving Eq. (1) consists in using the modal basis of the classically damped system yielding *M*=*P*+*S* equations of the form

$$\mathcal{M}_m \ddot{y}_m + \mathcal{C}_m \dot{y}_m + \mathcal{K}_m y_m = \boldsymbol{\phi}_m^T \mathbf{f}^e \quad \text{where} \quad \mathbf{u} = \sum_{m=1}^{M} \phi_m y_m \tag{2}$$

that can be solved readily. Assuming classical damping – which means that the damping matrix is diagonal in the modal basis – modal analysis is performed with the undamped system because the modal basis of the undamped system coincides with those of the damped system [2]. Accordingly, the eigenvalues $\omega_m^2$ and eigenvectors $\boldsymbol{\phi}_m$ of the system are computed solving

$$\mathbf{K}\boldsymbol{\phi}_m - \omega_m^2 \mathbf{M}\boldsymbol{\phi}_m = \mathbf{0} \tag{3}$$

In Eq. (2) and (3), we used the orthogonality properties of the modal basis and the following notations:

$$\mathcal{M}_m = \boldsymbol{\phi}_m^T \mathbf{M}\boldsymbol{\phi}_m \; ; \; \mathcal{C}_m = \boldsymbol{\phi}_m^T \mathbf{C}\boldsymbol{\phi}_m \; ; \; \mathcal{K}_m = \boldsymbol{\phi}_m^T \mathbf{K}\boldsymbol{\phi}_m \; ; \omega_m^2 = \mathcal{K}_m / \mathcal{M}_m \tag{4}$$

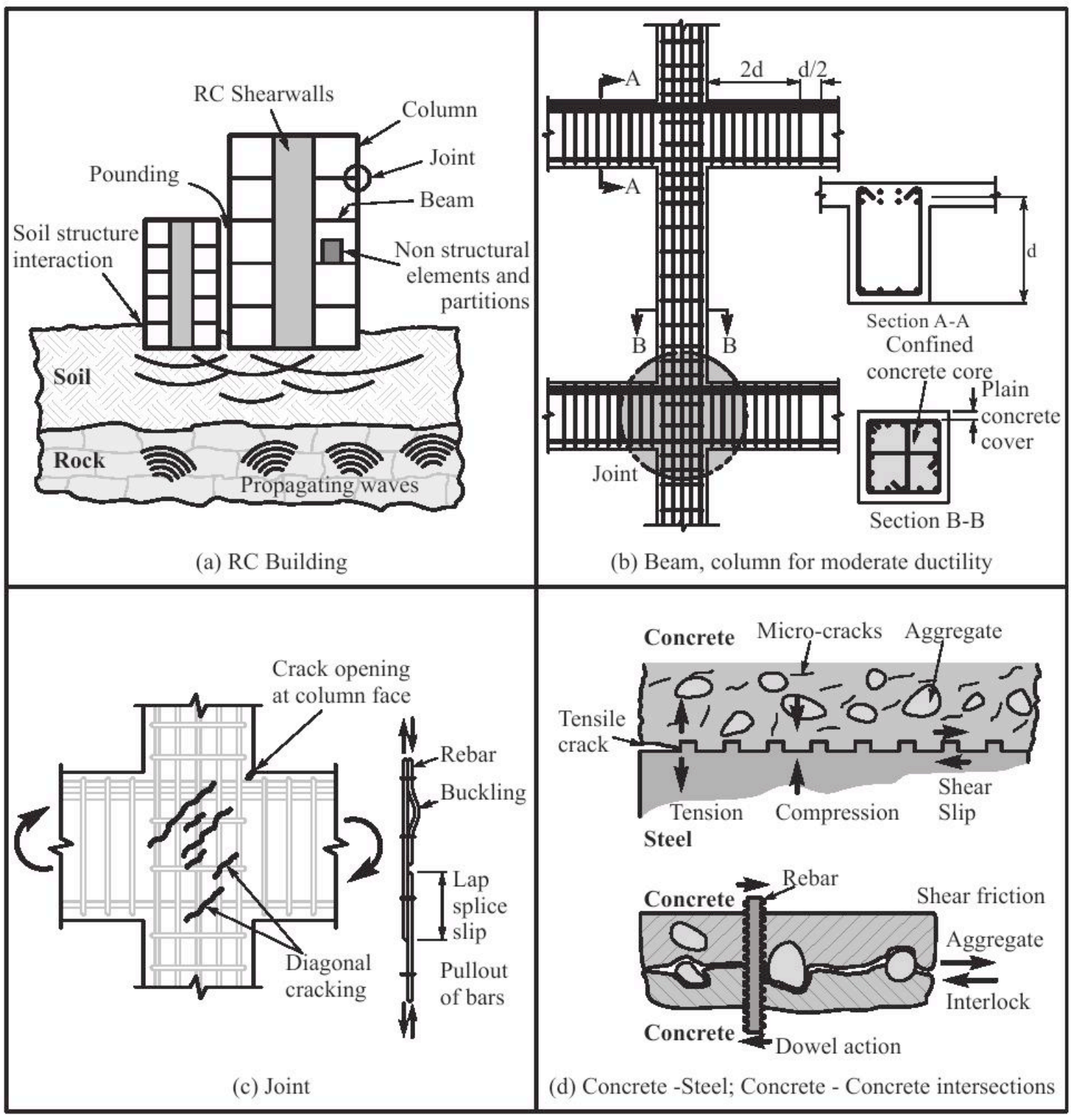

Fig. 1 – Some of the numerous energy dissipation mechanisms potentially activated in RC buildings in seismic motion. The hysteresis in the force-displacement curves used in nonlinear structural engineering results from a series of nonlinear phenomena from material to structural scales.

There are several options available to practitioners to build a viscous damping matrix **C** for seismic inelastic RHA. The consequences of using damping models initially developed for elastic systems in nonlinear RHA has been attracting the attention of researchers and practitioners for several decades (see [3] for a recent thorough review), and still is a topic of intense discussions (see the recent papers and discussion [3, 4, 5, 6]). The purpose of this paper is to provide practitioners with further insight into spurious damping forces that can be generated in nonlinear RHA. The term 'spurious' is used to refer to damping forces that are not present in an elastic system and develop as nonlinearities develop: such damping forces are not necessarily intended and appear as dependent on the structural history. In the following, two types of spurious damping forces are pointed out. Each type has often been treated separately in the literature, but each has been qualified as 'spurious', somehow blurring their differences. Then, in an effort to clarify the consequences of choosing a particular viscous damping model for nonlinear RHA, this paper shows that damping models that avoid spurious damping forces of one type do not necessarily avoid damping forces of the other type.

Next section reviews popular damping models that have been proposed for seismic nonlinear response history analysis: Caughey and Rayleigh damping along with Wilson-Penzien damping. In Section 3, two different types of spurious damping forces are presented and some strategies available for avoiding them – or controlling them – are discussed. Applications in Section 4 illustrate how viscous damping models that avoid one type of spurious forces still have the potential to generate spurious forces of another type. Finally, the issue of whether spurious viscous damping forces should really be avoided is raised and some conclusions with future directions toward a rational modeling of damping in inelastic RHA are formulated.

## 2. Models of viscous damping forces

Several damping models have been developed over the past decades for earthquake engineering. Rayleigh [7], Caughey [2, 8], and Wilson-Penzien [9] damping have been widely discussed in the literature. As Rayleigh damping is a particular Caughey series, we consider both together hereafter. All these three categories of damping models have been developed as nonlinear structural seismic analysis was very rare or even inexistent. In the context of the elastic analysis of multi-DOF structures, the main motivation was deriving damping matrices that would satisfy the hypothesis of classical – or proportional – damping.

## 2.1. Caughey and Rayleigh damping

The classical viscous damping model presented in Caughey's work for elastic structures [2, 8] is based on the following damping matrix:

$$\mathbf{C}^C(t) = \mathbf{M} \sum_{j=J_0}^{J_1-1} a_j(t) \left(\mathbf{M}^{-1}\mathbf{K}(t)\right)^j \tag{5}$$

Any $J$ significant structural eigenmodes, out of the total number of modes $M$, are considered in this relation. Stiffness matrix $\mathbf{K}$ is either defined as the initial stiffness matrix of the undamaged structure, or as the tangent stiffness. Besides, reduced forms of stiffness matrix can be used where for instance, as it will be discussed later, the components that pertain to massless DOFs can be set to zero.

As $J_0 = 0$ and $J_1 = 2$ in Eq. (1), Rayleigh damping matrix is retrieved:

$$\mathbf{C}^R(t) = a_0(t)\mathbf{M} + a_1(t)\mathbf{K}(t) \tag{6}$$

Viscous damping models of the Rayleigh category are pervasive in the Earthquake Engineering community. This category of damping models is rooted in the work of Lord Rayleigh [7] for elastic structural systems. It has been introduced as a mathematically convenient ad hoc procedure for representing damping.

Several kinds of Caughey or Rayleigh damping matrices have been implemented for nonlinear RHA depending on which stiffness matrix $\mathbf{K}$ is used (initial, tangent, reduced) and on whether the coefficients $a_0$ and $a_1$ are calculated once for all from initial structural properties or are updated throughout RHA. Theoretically, these damping coefficients can be computed for any $J$ of the $M$ structural eignemodes and at any time $t$ in the structure history.

## 2.2. Wilson-Penzien damping

Wilson and Penzien [9] have proposed the following damping matrix that yields classical damping in linear systems:

$$\mathbf{C}^{WP}(t) = \mathbf{M} \left( \sum_{j=1}^{J} \frac{2\xi_j(t)\omega_j(t)}{\mathcal{M}_j} \boldsymbol{\phi}_j(t)\boldsymbol{\phi}_j^T(t) \right) \mathbf{M} \tag{7}$$

This damping matrix does not benefit from the diagonal pattern of mass and stiffness matrices with the finite element procedure and has therefore not been popular in practice for computational reasons. However, it has been shown in [4] that, in the context of nonlinear RHA, the computational demand of this model is comparable with those of tangent stiffness-based Rayleigh damping. Besides, it has been shown in [3] that this model can be efficiently used when the structural damping matrix is built from the assembly of elemental damping matrices.

# 3. Spurious damping forces

The damping models presented above have been developed in the context of elastic structural analysis. As seismic nonlinear RHA emerged, the same damping models have been used in the simulations. However in nonlinear RHA, it has been observed on numerous occasions for several decades that these damping models can generate unintended spurious damping forces, in the sense that they can generate damping forces in the nonlinear regime that are not generated while the system remains elastic.

## 3.1. Spurious damping forces of type *S*

As Caughey or Rayleigh viscous damping model is used, there is no damping forces generated at the $S$ massless DOFs in elastic structures. However, as the structure yields, damping forces appear at these $S$ DOFs, and such forces have been qualified as 'spurious' because they develop in consequence of the yielding of structural elements, which seems anomalous [4]. This has first been pointed out in [10] and it has then been given a rational explanation in [11], which we briefly recall here.

Starting from Eq. (2), damping forces read

$$\mathbf{f}^d(t) = \mathbf{C}(t)\dot{\mathbf{u}}(t) = \sum_{m=1}^{M} \mathbf{f}^d_m(t) \qquad \text{with} \qquad \mathbf{f}^d_m(t) = \dot{y}_m(t)\mathbf{C}(t)\,\boldsymbol{\phi}_m(t) \tag{8}$$

And inertia forces read

$$\mathbf{f}^i(t) = \mathbf{M}\ddot{\mathbf{u}}(t) = \sum_{m=1}^{M} \mathbf{f}^i_m(t) \qquad \text{with} \qquad \mathbf{f}^i_m(t) = \ddot{y}_m(t)\mathbf{M}\,\boldsymbol{\phi}_m(t) \tag{9}$$

Also, from Eq. (3), the following relation holds

$$\left(\mathbf{M}^{-1}\mathbf{K}(t)\right)^j \boldsymbol{\phi}_m(t) = \omega_m^{2j}(t)\,\boldsymbol{\phi}_m(t) \tag{10}$$

which, introduced in Eq. (5) and using Eqs. (8) and (9), yields

$$\mathbf{f}^d_m(t) = \gamma_m(t)\,\mathbf{f}^i_m(t) \qquad \text{with} \qquad \gamma_m(t) = \frac{\dot{y}_m(t)}{\ddot{y}_m(t)} \sum_{j=0}^{J-1} a_j(t)\omega_m^{2j}(t) \tag{11}$$

This means that modal damping and modal inertia forces have the same shape. Accordingly, because $\mathbf{f}^i_{m,s} = \mathbf{0}$ for all modes $m$ at massless DOFs, there is also $\mathbf{f}^d_{m,s} = \mathbf{0}$ and eventually no damping forces generated at any of the $S$ massless DOFs.

In the same paper [11], it is also shown that

$$\mathbf{C}_{ss}(t) = \mathbf{0} \quad \Rightarrow \quad \mathbf{f}^d_s(t) = \mathbf{0} \tag{12}$$

which means that there is no spurious damping forces of type $S$ for any damping matrix that satisfies $\mathbf{C}_{ss}(t) = \mathbf{0}$ at any time $t$ throughout the analysis. Keeping this in mind, and remembering that $\mathbf{M}_{ps} = \mathbf{M}_{sp} = \mathbf{M}_{ss} = \mathbf{0}$, it is clear that Caughey damping with $J_1 < 2$ (see Eq. (5)) as well as Wilson-Penzien damping (see Eq. (7)) do not produce spurious forces of type $S$.

According to Eq. (12), another straightforward approach for avoiding damping forces of type $S$ consists in assigning zeros to the components of the damping matrix that pertain to a massless DOF, thus building a reduced stiffness matrix that satisfies the condition $\mathbf{C}_{ss}(t) = \mathbf{0}$ at any time $t$ throughout the analysis. This can be done either with static condensation (with zero loading at $S$ massless DOFs) [11] or without [12].

Still, of course, if stiffness is null after yielding, using the tangent stiffness for building the damping matrix also guarantees there is no such spurious damping forces [12]. In [13] spurious damping forces of type $S$ are avoided using elastic beam elements with zero-length semi-rigid rotational plastic hinges at their ends and assigning zero stiffness to the $S$ massless DOFs when building Rayleigh damping matrix.

### 3.2. Spurious damping forces of type *P*

The work presented in [14] has pointed out that another kind of spurious damping forces can appear in seismic nonlinear RHA as the natural frequencies of the structure drop in consequence of yielding or damage. This shift of the natural dynamic properties of the structure can result in a shift of the effective viscous damping ratios and consequently of the damping forces in the system. Following the assumption that viscous damping ratios should remain the same throughout inelastic RHA (as discussed in [4]), that is in the elastic range as well as after yielding or damaging, this shift of damping forces can be qualified as spurious. This second kind of spurious damping forces can affect both mass and massless DOFs depending on the damping model that is selected. We refer to them as spurious damping forces of type $P$.

For both Caughey and Wilson-Penzien damping models, spurious damping forces of type $P$ are generated in nonlinear RHA if the damping coefficients are set once for all, say at time $t_0$, while the structural modal properties change as the structure becomes nonlinear. The time history of the effective viscous damping ratios has been analytically derived in [15] for Rayleigh damping. This has also been investigated in [16]. A basic assumption in the formula derived in [15] is that the off-diagonal terms in the modal damping matrix can be neglected. Indeed, as the structure yields at time $t_y$, the matrix $\boldsymbol{\phi}^T(t_y)\mathbf{C}(t_0)\boldsymbol{\phi}(t_y)$ is not necessarily diagonal because $\mathbf{C}(t_0)$ is built from modal properties that are not computed at the same time $t_y$. We also use this assumption in what follows and therefore express effective modal viscous damping ratios time history from the relation:

$$\boldsymbol{\phi}_m^T(t)\mathbf{C}\boldsymbol{\phi}_m(t) = 2\,\xi_m(t)\omega_m(t)\boldsymbol{\phi}_m^T(t)\mathbf{M}\boldsymbol{\phi}_m(t) \tag{13}$$

We focus on Caughey series with negative powers and on Wilson-Penzien damping. It has been shown that those two models avoid spurious damping forces of type *S* [4, 11], but it is shown now that they do not necessarily avoid spurious damping forces of type *P*:

1) Introducing Eq. (5) with negative powers into Eq. (13), it comes ($J \geq 1$)

$$\xi_m(t) = \frac{1}{2}\sum_{j=0}^{J-1} a_j(t_0)\,\omega_m^{-2j-1}(t)\,h_m^{-j}(t) \quad \text{with} \quad h_m(t) = \frac{\boldsymbol{\phi}_m^T(t)\mathbf{K}_0\boldsymbol{\phi}_m(t)}{\mathcal{K}_m(t)} \tag{14}$$

Therefore, if the coefficients $a_j$ are calculated once for all, say at time $t_0$, the modal viscous damping ratios change according to the evolution of the modal properties.

2) Introducing now Eq. (7) into Eq. (13), we have

$$\xi_m(t) = \frac{1}{\mathcal{M}_m(t)\omega_m(t)}\sum_{j=1}^{J}\frac{\xi_j(t_0)\omega_j(t_0)}{\mathcal{M}_j(t_0)}\mathcal{M}_{mj}^2(t) \quad \text{with} \quad \mathcal{M}_{mj}(t) = \boldsymbol{\phi}_m^T(t)\mathbf{M}\boldsymbol{\phi}_j(t_0) \tag{15}$$

Here again, it is clear that the modal damping ratios change with the modal properties as nonlinearity appears.

For both models, as shown in Section 3.1, there is no damping forces generated at *S* DOFs. However, spurious damping forces are introduced at *P* DOFs if we adopt the assumption that modal viscous damping ratios should remain unchanged in the elastic and nonlinear regimes. With this assumption, any damping force that would appear in the nonlinear regime while absent in the elastic one is qualified as spurious.

Solutions to avoid spurious damping forces of type *P* have been proposed in the past. In [14], Rayleigh damping is introduced in nonlinear RHA with coefficients updated at each time step, which leads to constant viscous damping ratios for the two modes that are considered for building Rayleigh damping model. In [17], a capped viscous damping model is presented based on the introduction of bounds for the effective viscous damping ratios throughout the nonlinear RHA. This notion of bounds has also been considered later in [15] with emphasis put on the consequences of using initial or tangent stiffness for Rayleigh damping: it is shown that there is no choice that is intrinsically better than the other but that it is easier to control these bounds when tangent stiffness is adopted. In [12], the author provides recommendations to manage the issues raised by spurious damping forces of both types *P* and *S*, and ultimately comes up with the conclusion that viscous damping model should ideally be eliminated and replaced by hysteretic laws that would represent the actual sources of the overall structural damping.

Very recently, the approach that has been presented in [3] combines the benefits of Wilson-Penzien damping [4, 8] (control over a large range of modes) and updated damping coefficients [14] (control over the viscous damping ratios history), and solves the computational issue raised by the need to re-compute the modal basis at each time step. This approach is based on building a viscous damping matrix from the finite element assembly of elemental damping matrices (as for stiffness or mass matrices) instead of building a global (structural) damping matrix. This allows for using analytical expression for solving the eigenvalue problem at the element level.

## 4. Applications

The applications below show that damping models that avoid spurious forces of type *S* do not necessarily avoid spurious damping forces of type *P*.

## 4.1. Inelastic structure and yielding scenarios

The structure used for this application is the same as the one introduced in [12], which we briefly present here. The structure is a five-story building modeled as a system of five DOFs – the horizontal displacements – connected by inelastic columns, which all have the same elastic properties. At each DOF the same mass is lumped. Two different yielding scenarios are considered:

1. The entire stiffness matrix is assumed to uniformly drop to 50% of its original value;
2. The structural elements non-uniformly yield along the building height: $N$th-story stiffness is reduced to $10\% + (N-1) \times 20\%$ of its original value, that is 10% for the 1st story, 30% for the 2nd… 90% for the 5th.

To investigate the history of the damping ratios while the structure is yielding, we arbitrarily set the total duration of the inelastic RHA to $T = 1$ s and divide the analysis into 5 steps (0, 0.2 s, . . . 1 s). At each time step the structure suddenly damages. Here is the damaging history for scenario 1 (uniform damaging): at $t = 0$ structural stiffness matrix $\mathbf{K}$ is equal to the initial stiffness matrix $\mathbf{K}_0$, then at $t = 0.2$ s it suddenly drops to $\mathbf{K} = 90\% \times \mathbf{K}_0$, at $t = 0.4$ s to $\mathbf{K} = 80\% \times \mathbf{K}_0$, and so on until $t = 1$ s where $\mathbf{K} = 50\% \times \mathbf{K}_0$.

The structure has 5 eigenmodes, with corresponding angular eigenfrequencies in the initial (undamaged) state: $\omega_1(t_0), \ldots, \omega_5(t_0) = 5.56, 16.23, 25.58, 32.87, 37.49$ rad/s.

## 4.2. Caughey and Rayleigh damping

The performance of initial and tangent stiffness-based Rayleigh damping with either frozen or updated coefficients, along with Rayleigh damping with reduced stiffness (with for instance zeros assigned to the components pertaining to the $S$ DOFs) has been investigated in [15]. Here we focus on the Caughey damping model proposed in [11] to solve the problem of spurious damping forces of type $S$:

$$\mathbf{C} = \mathbf{M}\sum_{j=0}^{1} a_j(\mathbf{M}^{-1}\mathbf{K}_0)^{-j} = a_0\mathbf{M} + a_1\mathbf{M}\mathbf{K}_0^{-1}\mathbf{M} \tag{16}$$

This model corresponds to Caughey series (Eq. (5)) with $J_0 = J_1 = 0$ (negative power) and is based on the initial stiffness matrix $\mathbf{K}_0$. Viscous damping ratios time histories are computed from Eq. (14) with $J = 2$. The coefficients $a_0$ and $a_1$ are computed from the initial structural properties ($t = t_0$) so that the viscous damping ratios for modes 1 and 3 are $\xi_1(t_0) = \xi_3(t_0) = \xi_0 = 2\%$; these coefficients are set once for all and kept frozen throughout the inelastic RHA. Angular eigenfrequencies $\omega_m(t)$ and factors $h_m(t)$ are computed at every time step of the nonlinear RHA ($t = 0.2$ s, …, $t = 1$ s).

The evolution of the effective modal viscous damping ratios $\xi_m(t)$ is plotted in Fig. 2. In case of uniform stiffness degradation (top row in Fig. 2), effective viscous damping ratio for mode 1 increases constantly from the initially set $\xi_1 = 2\%$ (at $t = 0$) to almost $\xi_1 = 6\%$ at the end of the RHA at $t = 1$ s. In case of non-uniform stiffness degradation (bottom row in Fig. 2), this increase is even larger, with an effective viscous damping ratio for mode 1 getting larger than 12% at the end of the RHA. These large increases are due to the shift of the modal frequencies because of structural yielding. It is therefore obvious here that, although the damping model described in Eq. (16) does not generate spurious damping forces of type $S$ as discussed earlier, it can generate spurious damping forces of type $P$.

## 4.3. Wilson-Penzien damping

We now focus on Wilson-Penzien damping model. The damping matrix $\mathbf{C}^{WP}$ (see Eq. (7)) is built from the initial structural properties ($t = t_0$) with $J = 5$. The viscous damping ratios are set to $\xi_j(t_0) = \xi_0 = 2\%$ for all 5 modes $j = 1, \ldots, 5$. The modal structural properties are computed at each of the 5 time steps considered in the inelastic RHA and the corresponding time history of the effective viscous damping ratios is computed using Eq. (15).

The evolution of the effective modal viscous damping ratios is plotted in Fig. 3. In the case of uniform stiffness degradation (top row in Fig. 3), the damping ratios constantly and uniformly increase from $\xi_1(t{=}0) =$ $\xi_0$ = 2% to more than $\xi$(t=1s) = 2.8% as the structure yields. In the case of non-uniform stiffness degradation (bottom row in Fig. 3), the effective viscous damping ratio for mode 1 constantly increases from $\xi_1$(t=0) = $\xi_0$ = 2% to $\xi_1$(t=1s) = 5.2%. As for Caughey or Rayleigh damping, it is obvious that, although the damping model described in Eq. (7) does not generate spurious damping forces of type *S* as discussed above, it can generate spurious damping forces of type *P*.

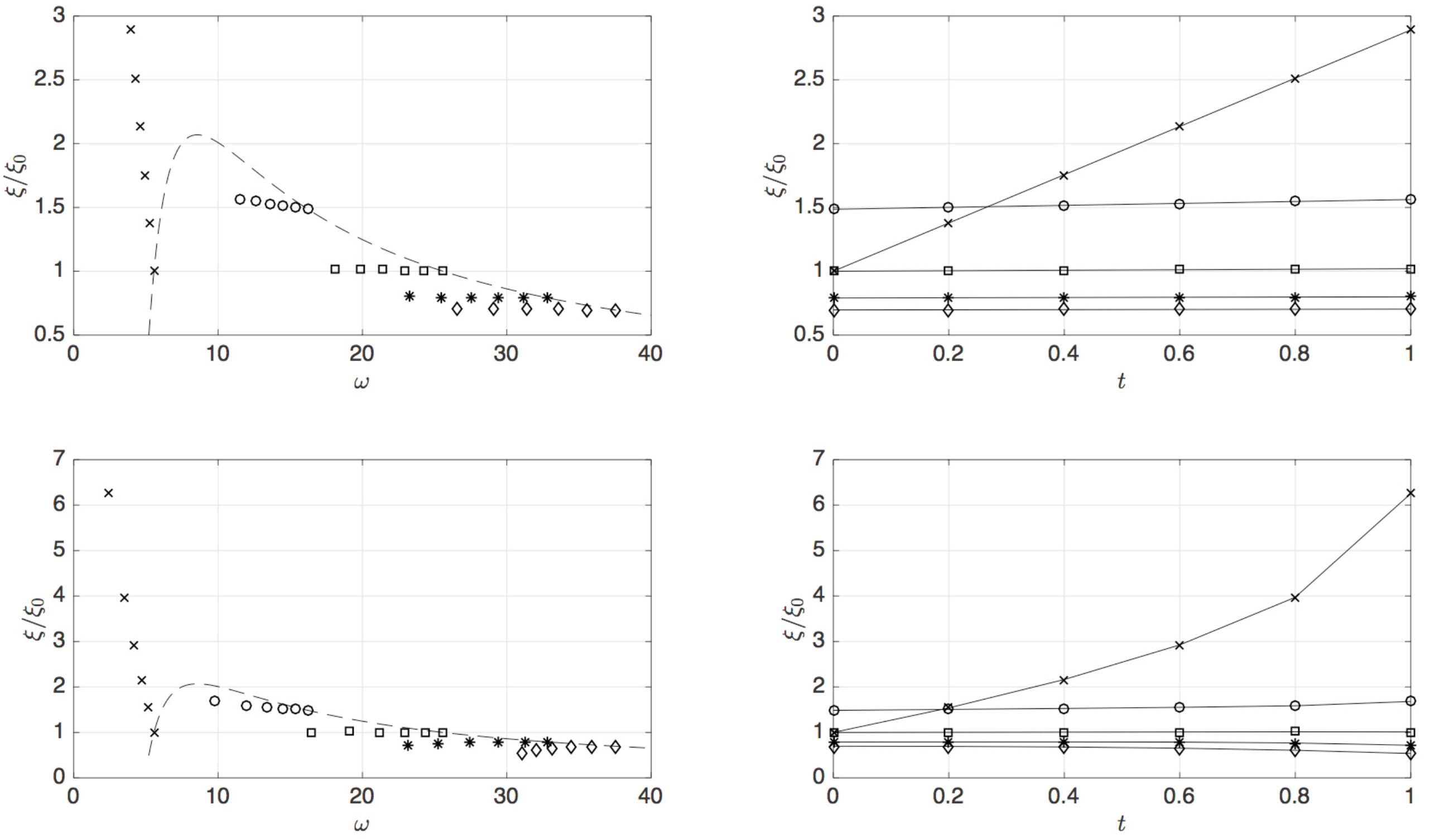

Fig. 2 – Viscous damping ratios time history for modes 1 (×), 2 (o), 3 (□), 4 (★) and 5 (◇) with damping matrix as in Eq. (16) in case of uniform [top] and non-uniform [bottom] structural yielding scenarios. Damping coefficients $a_0$ and $a_1$ are identified once for all according to the initial elastic structural properties (at time t = 0) so that initial damping ratios $\xi_1 = \xi_3 = \xi_0$ = 2%. Dashed lines represent the curve $\xi(\omega;\, t = 0)$.

## 5. Should we really avoid ‘spurious’ damping forces?

As mentioned on several occasions in this paper, the term ‘spurious’ has been used to qualify damping forces that are triggered by the yielding of the structure and that are not present as the structure remains in its initial elastic domain. However, as the structure yields, it is fair to recognize that the numerous energy dissipative mechanisms that can be activated in yielding or damaging structural systems (see Fig. 1) are not all well understood, let alone well modeled. Consequently, additional damping forces generated in inelastic incursions could be seen as physical, for instance in the case where nonstructural elements – which are generally not modeled in the hysteretic structural response – participate to the overall damping in the structure.

On another hand, there is no evidence that the energy dissipative mechanisms that generate overall damping in the elastic regime remain unchanged in the nonlinear regime, although this is a tacit assumption in nonlinear RHA [4]. Nevertheless, as far as spurious damping forces of type *P* are concerned, this is because of this assumption that damping forces are referred to as ‘spurious’ as an increase of the effective modal viscous damping ratios is observed.

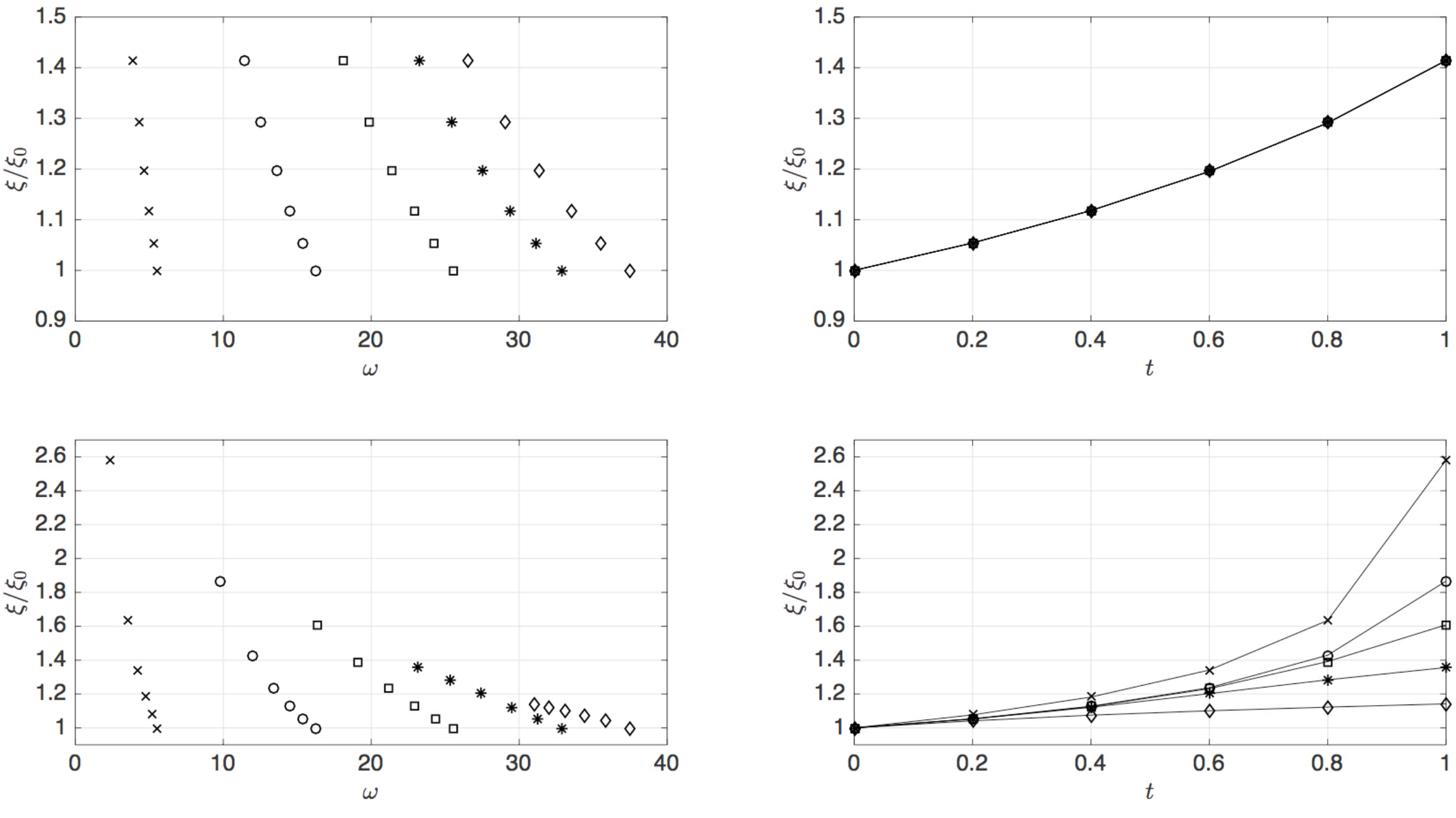

Fig. 3 – Viscous damping ratios time history for modes 1 (×), 2 (o), 3 ( ), 4 (★) and 5 (◊) with damping matrix as in Eq. (7) in case of uniform [top] and non-uniform [bottom] structural yielding scenarios. Initial damping ratios (at time t = 0) are set to $\xi_j = \xi_0 = 2\%$ for all modes $j = 1,\ldots, 5$.

## 6. Conclusions

In this paper, two types of 'spurious' damping forces have been identified in nonlinear seismic response history analysis (RHA). 'Spurious' damping forces are forces that arise – possibly unintended – during inclusions in the nonlinear regime and that are absent in the elastic domain. 'Spurious' damping forces of type *S* develop at massless DOFs as the very pervasive Rayleigh damping model is used, as well as for any other Caughey series. 'Spurious' damping forces of type *P* develop as the structural modal frequencies drop in consequence of yielding or damaging, whether Rayleigh or Caughey or Wilson-Penzien damping is used.

According to the state of the practice in the field of seismic inelastic RHA, it is recommended to choose viscous damping models that avoid both types of spurious damping forces. However, as pointed out in Section 5, whether so-called spurious viscous damping forces should be avoided or not is not that straightforward because the fact is that the actual damping forces in nonlinear structural systems are not always well identified nor modeled. Nevertheless, it is of utter importance that what is actually modeled in nonlinear RHA be controlled and it is therefore worth gaining better knowledge on the damping forces that are actually generated when using this or that viscous damping model. This is what this paper intends to provide.

Toward the purpose of designing reliable well-controlled damping models, next step to proceed to is, on the one hand, the quantification of the viscous damping forces that are actually modeled and, on the other hand, the identification of the damping forces that are actually needed. The concept of discrepancy forces introduced in [18] mingles structural simulation and experimental data, which allows quantifying the actual damping forces needed to satisfy equilibrium. From these discrepancy forces, damping models could be identified on a rational basis: this is what our ongoing investigations on this topic are oriented toward.

## 6. Acknowledgements

This research has been partly supported by a Marie Curie International Outgoing Fellowship within the 7th European Community Framework Programme (proposal No. 275928).